\documentclass[12pt]{article}

\usepackage{dsfont}
\usepackage[english]{babel}
\usepackage[utf8]{inputenc}
\usepackage{amsmath}
\usepackage{graphicx}
\usepackage[colorinlistoftodos]{todonotes}
\usepackage{hyperref}
\usepackage{amsfonts}
\usepackage{subfigure}
\usepackage{bm}
\usepackage[toc,page]{appendix}
\usepackage{natbib}
\usepackage{geometry}
\usepackage{setspace}
\newcommand{\bcdot}{\boldsymbol{\cdot}}

\definecolor{Lblue}{rgb}{0.13, 0.36, 0.51}

\newcommand\blfootnote[1]{%
  \begingroup
  \renewcommand\thefootnote{}\footnote{#1}%
  \addtocounter{footnote}{-1}%
  \endgroup
}

\title{Measuring Effects of Medication Adherence on Time-Varying Health Outcomes using Bayesian Dynamic Linear Models\blfootnote{Work supported by NIH grant R21 HL121366}}
\author{Luis F. Campos, Mark E. Glickman, Kristen B. Hunter}
\date{}

\begin{document}
\maketitle
\begin{abstract}
\begin{spacing}{1}
One of the most significant barriers to medication treatment is patients' non-adherence 
to a prescribed medication regimen.
The extent of the impact of poor adherence on resulting health measures is often unknown,
and typical analyses ignore the time-varying nature of adherence.
This paper develops a modeling framework for longitudinally recorded health measures modeled
as a function of time-varying medication adherence 
or other time-varying covariates.
Our framework, which relies on normal Bayesian dynamic linear models (DLMs), 
accounts for time-varying 
covariates such as adherence and non-dynamic covariates such as baseline health 
characteristics.
Given the inefficiencies using standard inferential procedures for DLMs
associated with infrequent and irregularly recorded response data,
we develop an approach that relies on factoring the posterior density into
a product of two terms;
a marginal posterior density for the non-dynamic parameters,
and a multivariate normal posterior density of the dynamic parameters conditional
on the non-dynamic ones.
This factorization leads to a two-stage process for inference in which
the non-dynamic parameters can be inferred separately from the time-varying parameters.
We demonstrate the application of this model to 
the time-varying effect of anti-hypertensive medication on blood pressure levels from
a cohort of patients diagnosed with hypertension.
Our model results are compared to ones in which adherence is incorporated through
non-dynamic summaries.
\end{spacing}
\end{abstract}


\section{Introduction} 
\label{sec:introduction}
Over 85 million American adults, or about one third of the population over 20 years old, 
suffer from hypertension 
\citep{benjamin2018heart}.
Approximately 16\% of adults in the United States 
are unaware that they have hypertension
\citep{benjamin2018heart}.
Left untreated, hypertension can lead to a range of serious and costly health concerns
such as cardiovascular disease, stroke, and renal disease 
\citep{amery_mortality_1985, probstfield_prevention_1991}.
Among the many factors associated with uncontrolled BP, poor adherence to prescribed 
anti-hypertensive medications is of increasing concern to clinicians, health care systems, 
and other stakeholders 
\citep{choo_validation_1999, morisky_concurrent_1986, osterberg_adherence_2005}. 
Little doubt exists that patients who adhere poorly to their prescribed medication are 
at risk for worse BP outcomes. 
Given the large variation in adherence, both across patients and temporally within 
patients, it is an open question how to accurately measure the impact of
varying adherence patterns on BP levels.
Furthermore, the variation in adherence patterns creates difficulties in accurately
measuring the effects of socio-demographic and health risk factors on BP levels.

This paper develops a Bayesian dynamic linear model 
\citep{durbin_time_2001,west_bayesian_1997,Petris:2009cs} 
for health measures recorded over time 
as a function of time-varying adherence, with a particular
application to the effects of anti-hypertensive medication on BP levels.
Bayesian dynamic linear models 
(DLMs) have a long history as a statistical framework for forecasting and measuring 
trajectories in many domains, including real-time missile tracking and financial 
securities forecasting, but rarely in the health context. 
We apply DLMs to describe time-varying health measures (like blood pressure levels)
as a function of detailed adherence (or other time-varying covariates) and individual 
demographics and comorbidities measured typically at study baseline. 
The application of DLMs to the time-varying adherence on health measures is novel, but 
fits naturally into the DLM framework because these measures can be ``tracked'' over 
time as adherence data accumulate.  
Because the DLM framework permits the inclusion of patient-level predictors, 
the model can be applied to measuring effects of socio-economic or racial disparities, 
or the effects of different comorbid conditions.
In these settings, our model can control for differential medication adherence patterns, 
resulting in more accurate measurement of the effects of other covariates. 

Data for studies in which the estimated effects of time-varying adherence on health measures 
are desired usually consist of three components.
First, adherence data for each study participant 
are assumed to be collected through electronic monitoring devices.
These devices electronically time-stamp each time the pill container is opened,
and permit an accurate recording of when a patient took their medication.
Second, health and socio-demographic are typically recorded at the start of
medication adherence studies, and are often non-dynamic.
Finally, health measures which might be impacted by differential medication adherence
are recorded over time at intermittent intervals.
Such measures are often recorded at clinic visits, the timing of which may be determined
by the patient.
Thus it is quite common for the number of health measures per study patient
to be much fewer than the number of days in which medication adherence
information is recorded.

The DLM framework is challenged by having much fewer days with health measures
compared to the number of days of adherence measurements.
In typical uses of DLMs, both time-varying covariates and responses are measured
at every time point, and inference for the time-varying state parameters can be 
accomplished using standard Bayesian updating algorithms \citep[Chapter~16]{west_bayesian_1997}.
With adherence data, when the responses are measured at irregular and infrequent intervals,
the usual updating approaches can be demonstrably inefficient.
Instead, we develop an inferential approach in a Bayesian setting that takes advantage of 
factoring the posterior density into a product of two terms.
The first term involves
the exact marginal likelihood of the DLM, marginalizing out the dynamic state process. 
The second is the conditional posterior density for the state process parameters. 
These together allow us to determine the posterior distribution for the non-dynamic parameters 
using standard Markov chain Monte Carlo (MCMC) techniques as a first stage and the state process 
parameters as a second stage without resorting to needlessly complex computational tools. 
With a DLM that has normally distributed responses and a stochastic process for the latent
states the has normal innovations, the factorization is the product of two multivariate
normal densities.
This factored posterior density also easily permits inference for the non-dynamic 
parameters, which in the setting of medication adherence is likely of interest because 
the researcher may want to understand the effect of baseline health characteristics
on the health measures controlling for differential adherence.

The remainder of this paper is organized as follows. 
In Section \ref{sec:data} we introduce a motivating example and details of the study cohort.
We specify the DLM for our framework in Section \ref{sec:methods}.  
The model, which assumes an autoregressive structure, 
accounts for possibly multivariate health measures which may or may
not be measured simultaneously.
We then develop our computational approach for inference in 
Section \ref{sec:marginal_dynamic_linear_models}.  
We apply our methods to modeling BP in Section \ref{sec:application_to_bp_study} where 
we compare our methods to typical models used to measure the effects of adherence. 
We conclude in Section~\ref{sec:discussion} with a discussion on the limitations and 
potential extensions of our methods.

\section{Data} 
\label{sec:data}
The data we analyzed were obtained from the baseline pre-randomization period of a trial that studied the effects of a provider-patient communication skill-building intervention on adherence to anti-hypertensive medication and on BP control (clinicaltrials.gov ID:  NCT00201149).  Patients were recruited from seven outpatient primary care clinics at Boston Medical Center, an inner-city safety-net hospital affiliated with the Boston University School of Medicine.  Patients enrolled into the study from August 2004 and June 2006, meeting several eligibility criteria.  These included that the patients were of white or black race (African or Caribbean descent), were at least 21 years old, had an outpatient diagnosis of hypertension on at least three different occasions prior to study enrollment, and were currently on anti-hypertensive medication.  The cohort size was 869 patients. The study involved measuring anti-hypertensive medication-taking using Medication Event Monitoring System (MEMS) caps, a particular type of electronic pill-top monitoring device.  Patients were given their most frequently taken anti-hypertensive medications in a bottle with a MEMS cap, and were instructed to open the bottle each time they took a dose.  
Each MEMS cap contained a microprocessor that recorded the date and time whenever the bottle was opened, and the timing information was then downloaded through a wireless receiver after the patient returned the MEMS cap.  Our study focused on medication-taking behavior during the entire pre-randomization baseline period of the study.  
A patient was considered adherent to the prescribed medication on a day if the MEMS cap was opened as many times as the daily medication dosing frequency, and not adherent otherwise. 
These measurements were recoded on a daily basis, but 
the duration over which adherence information was measured varied by patient.
Blood pressure measurements were recorded less 
frequently as they were obtained as part of routine clinical care.  

The BP readings were assessed using manual or electronic devices, and were obtained by clinical staff including physicians, nurses and medical assistants.  
In cases where multiple readings were obtained on a single day (typically at the same clinic visit), the individual values were recorded.  
Diastolic and systolic blood pressure values (DBP and SBP respectively) were recorded separately. 

In addition to detailed time-varying adherence and BP readings, other 
patient-specific baseline information was collected.
Race (white versus African American), gender, income, and age at the start of the study were recorded for each patient. 
From electronic medical records, the following comorbidities (as binary variables) were investigated in our study, given their potential impact on overall BP levels:  
Presence of cerebrovascular disease, congestive heart failure, chronic kidney disease, coronary artery disease, diabetes mellitus, hyperlipidemia, peripheral vascular disease, and obesity (defined as body mass index greater than $30 kg/m^2$).

\begin{figure}[ht!]
    \begin{center}
\includegraphics[scale=0.6]{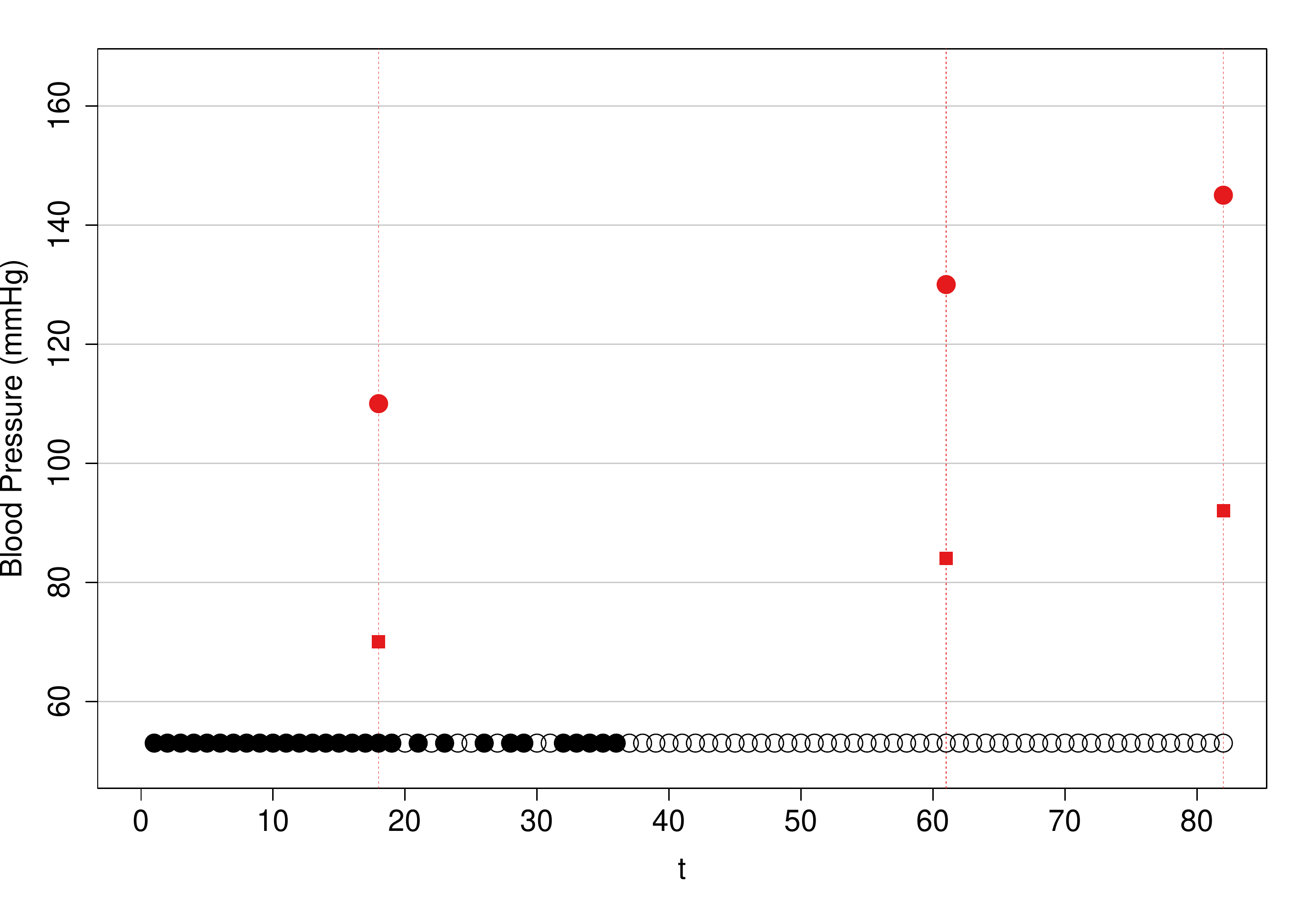}
    \caption{Example BP and adherence data for a study patient. 
    Red squares are diastolic BP, red dots are systolic BP, and
    the black dots near the horizontal axis indicate whether the patient
    was adherent on a particular day (filled dots indicate adherence,
    open dots indicate non-adherence).
    \label{fig:data_example}
    }
    \end{center}
\end{figure}

To motivate our modeling framework, consider the data 
from one of the patients in the study cohort
displayed in
Figure \ref{fig:data_example}. 
The figure shows the DBP and SBP measures on the three days the patient 
had their blood pressure measured,
and daily indications of whether the patient was adherent.
This patient was adherent
for a total of $29$ of $82$ days. 
However, this simple summary masks the time-varying pattern of the
patient's adherence.
The patient began the study by being fully adherent
to their prescribed medication, during which time their blood pressure
remained under control.
After 20~days, the patient started becoming less adherent,
and starting around day 38 the patient discontinued taking their
medication altogether.
Over this latter period, the patient's BP increased, and by the end
of the study period the patient has DBP and SBP values that were not under control.
Besides suggesting a potential link between adherence and BP measures, this 
example motivates a dynamic linear model for health measures that accounts 
for time-varying adherence.


\section{A dynamic model for multivariate health measures} 
\label{sec:methods}
We propose a statistical framework for multivariate time-varying health measures as a 
function of medication adherence that is a member of the class of Bayesian dynamic linear models. 
Let $y_{itk}$ be the value of the $k^{th}$ health measure, $k=1,\ldots,K$, 
for patient $i$ at time $t$, $t=1,\ldots,T_i$.
We assume that time is discretized and equally spaced 
(e.g., days), and that health measures are observed only at times
$t_1, t_2,\ldots, t_{m_i}$.
Let $\bm{y}_{it\bcdot}$, generally, indicate a vector collecting all outcomes 
across the dotted index into a column vector. 
In this example we collect the outcomes for patient $i$ at time $t$, 
i.e., $\bm{y}_{it\bcdot} = (\bm{y}_{it1}, ..., \bm{y}_{itK})^T$. 
Similarly, $\bm{y}_{i\bcdot k}$ collects the $k^{th}$ outcome observed 
at times $(t_1, ..., t_{m_i})$ for patient $i$, 
i.e., $\bm{y}_{i\bcdot k} = (\bm{y}_{i{t_1}k}, ..., \bm{y}_{i{t_m}k})^T$. 
Our framework assumes that $\bm{y}_{it\bcdot}$ is  an observed measurement 
generated from a distribution with mean
$\bm\mu_{it\bcdot}$ which follows a stochastic process. 
The framework recognizes that actual measurements on a given day could vary 
around a mean level due to various influences including emotional state, 
activity level, recent alcohol consumption, ambient temperature, and other 
unobserved factors that might affect the outcomes.  

The mean process at time $t$ is modeled as the sum of the contributions
of a non-dynamic term and a stochastic term.
Specifically, we assume
\begin{align}
\label{eq:DLM_static}
	\bm{y}_{it\bcdot}  &= \bm\mu_{it\bcdot}  + \bm\varepsilon_{it} \nonumber \\ 
	 &=  \bm\beta \bm{x}_i + \bm\alpha_{it\bcdot} + \bm\varepsilon_{it} 
\end{align}
where ${\bm{x}}_i$ is a vector of $p$ non-dynamic covariates typically
measured at baseline and $\bm\beta$ are the corresponding linear coefficients,
and where $\bm\alpha_{it\bcdot}$ is a stochastic process that may depend on
dynamic covariates, such as time-varying adherence.
Here $\bm{y}_{it\bcdot}, \bm\mu_{it\bcdot}, \bm\alpha_{it\bcdot}$ 
and $\bm\varepsilon_{it}$ are all vectors of length $K$ corresponding
to the observation, 
mean outcome process, latent state model and sampling error respectively. 
The non-dynamic covariate effects $\bm\beta$ are of dimension $K\times p$. 
We use $\bm\beta_k$ to denote the $k^{th}$ row of $\bm\beta$, i.e., 
the non-dynamic covariate effects on the $k^{th}$ outcome. 
The error term for subject $i$ at time $t$, $\bm\varepsilon_{it}$, 
accounts for typical sampling variability, measurement error and 
possible correlation of the outcomes within patient and time. 
We further assume that $\bm\varepsilon_{it} \sim \mbox{N}(0, \; \Sigma_\varepsilon)$
with possibly non-diagonal covariance matrix $\Sigma_\varepsilon$.

If patient $i$'s adherence is tracked for $T_i$ consecutive days, we 
indicate this measurement set as $\{1, ..., T_i\}$, 
and denote the collection of time-varying adherence measures as
$\{c_{it}\}_{t = 1}^{T_i}$. 
We let $c_{it} = 1$ if patient $i$ was adherent on day $t$ and $-1$ otherwise. 
As we describe below, we extend the definition of $c_{it}$ to include other time-varying
covariates.
We further let 
$\mathcal{T}_i = \{t_1, ..., t_{m_i}\} \subset \{1, 2, ..., T_i\}$
denote the set of $m_i$ days on which the outcomes were measured for patient $i$.
The set of outcome measurements for patient $i$ is denoted by
$\bm{y}_{i\bcdot\bcdot} = \{\bm{y}_{it\bcdot}\}_{t\in \mathcal{T}_i}$. 
Crucially, we assume that $m_i << T_i$ for all $i$.

Changes in 
$\bm\alpha_{it\bcdot}$ 
are assumed to be reflected by whether a patient takes the prescribed 
medication along with any other time-varying covariates.  
We assume an $AR(1)$ model on the latent states $\{\bm\alpha_{it\bcdot}\}$ and incorporate  the 
$\bm{c}_{it\bcdot}$ 
as time-varying covariates.
Other forms of stochastic processes are possible, including autoregressive processes 
of higher order.
More generally, if
$\bm{c}_{it\bcdot}$ 
is of dimension~$r$, then we assume a process on 
$\bm\alpha_{it\bcdot}$ 
given by
\begin{align}
	\label{eq:DLM_dynamic}
	\bm\alpha_{it\bcdot} &= \bm \rho \bm\alpha_{i,t-1,\bcdot}  + \bm\phi \bm{c}_{it\bcdot}  + \bm\nu_{it}
\end{align}
where $\bm \rho$ is the $K$-dimensional diagonal matrix of first-order autoregressive
parameters,  that is,
$\bm \rho = \text{diag}(\rho_1, \rho_2, ..., \rho_K)$. 
We assume $|\rho_k| < 1$ to ensure stationarity. 
The time-varying outcome effects $\bm \phi$ is a $K\times r$ matrix. 
We let $\bm\phi_k$ denote the $k^{th}$ row of $\bm\phi$, i.e., 
the time-varying covariate effects on the $k^{th}$ outcome. 
We further assume in~(\ref{eq:DLM_dynamic}) that the innovations $\bm\nu_{it}$ have 
zero-mean MVN distributions with diagonal covariance matrix 
$\Sigma_\nu = {\text{diag}}(\sigma^2_{\nu 1}, \sigma^2_{\nu 1}, ..., \sigma^2_{\nu K})$. 
Similarly we assume $\bm\alpha_{i1} \sim N_K(0,\Sigma_0)$ with 
$\Sigma_0 = {\text{diag}}(\sigma^2_{0 1}, \sigma^2_{0 1}, ..., \sigma^2_{0 K})$. 
Let $\theta$ denote the non-dynamic parameters 
$\theta = (\bm\beta, \Sigma_\varepsilon, \bm \rho, \bm\phi, \Sigma_\nu, \Sigma_0)$. 
Equations~(\ref{eq:DLM_static}) and~(\ref{eq:DLM_dynamic}) define the distributions 
$p({\bm{y}}_{it\bcdot} |\bm\alpha_{it\bcdot},\bm{x}_i, \theta)$ 
and $p(\bm\alpha_{it\bcdot} |\bm\alpha_{i,t-1,\bcdot}, \theta, \bm{c}_{it\bcdot})$ respectively.  

\begin{figure}[ht!!]
    \begin{center}
        \includegraphics[scale=0.6]{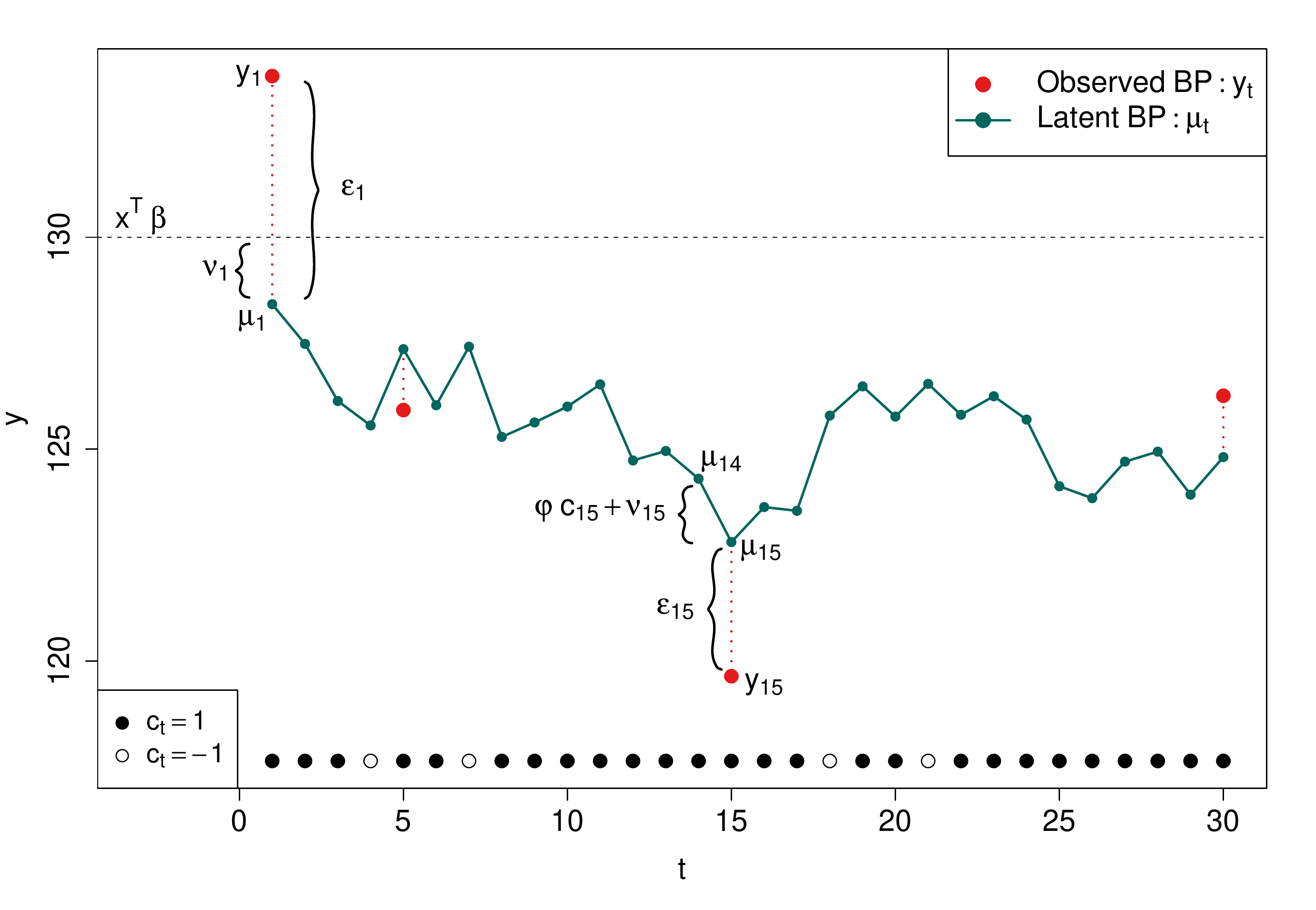}
    \end{center}
    \caption{
    \label{fig:ssm_cartoon}
    A 30-day trajectory for a simulated patient from a 
    DLM with a scalar outcome.}
\end{figure}

Figure \ref{fig:ssm_cartoon} contains a 
simulated example of a patient with a scalar outcome
over a 30-day period corresponding to the model in
(\ref{eq:DLM_static}) and (\ref{eq:DLM_dynamic}).
The bottom of the figure displays the adherence indicator
simulated independently with a 90\% probability of adherence. 
The adherence effect is simulated to be large and negative 
($\phi = -0.5$).
The contribution of the non-dynamic covariates is assumed to be 
$x_i^T\beta=130$. 

As evidenced in Figure~\ref{fig:ssm_cartoon}, the mean process generally 
declines on days when a patient is adherent.
This is not always the case, and an increase can occur when the corresponding
innovation $\nu_t$ is large and positive offsetting the impact
of the patient taking their medication.
The observed health measure is normally distributed around the
mean for the day on which the measure is recorded.
On day~1, for example, the health measure is higher than the 
mean, and on day~15 the health measure is lower than the mean.

This model is attractive for relating health measures to time-varying
adherence for several reasons. 
First, the model can account for non-dynamic baseline variables as 
well as time-varying adherence.
This model feature permits disentangling the effects of detailed
adherence from patient-specific socio-demographic and health 
covariates.
Second, in settings where multiple outcome measures are
observed at time $t$, inference for the sampling variability
$\Sigma_\varepsilon$
can be made more precise, and help to separate 
from the 
innovation variability $\Sigma_\nu$. 

The assumption of an $AR(1)$ model component has a well-known 
asymptotic mean under full adherence and full non-adherence.  
Specifically, 
the overall effect of repeated days of adherence on the outcome 
measure
converges for increasing values of $t$ as 
$\bm\alpha_{it\bcdot} \rightarrow (I-\bm\rho)^{-1}\bm\phi$.  
We can therefore predict that if patient $i$ were to continue to 
be fully adherent, their mean outcomes would tend to 
$\bm\mu_{it\bcdot}\rightarrow \bm\beta \bm x_i  + 
(I-\bm\rho)^{-1}\bm\phi$. 
An analogous calculation can be performed when the patient is 
fully non-adherent. 
This property permits estimating the best-case or worst-case 
health measure means for perfect adherence or perfect non-adherence,
even when patients' adherence level is somewhere in between.

\section{Marginal Dynamic Linear Models} 
\label{sec:marginal_dynamic_linear_models}

Inference for 
the model in Section~\ref{sec:methods} is challenging given the large number
of parameters, both the non-dynamic parameters $\theta$ as well as the time-varying
parameters ${\bm{\alpha}}$. 
In particular,
if the  number of recorded adherence indicators per patient is large,
then the size of ${\bm{\alpha}}$ is similarly large.
Standard Bayesian inferential methods through posterior simulation approaches
for such a highly parameterized model 
can result in slow convergence and unreliable inferences.

Advances in Bayesian computation have made highly parameterized 
dynamic models more tractable.
With recent developments in sequential Monte-Carlo (SMC) with software packages like 
Libbi \citep{2013arXiv1306.3277M}, sampling the induced high-dimensional latent states
of the DLM has become computationally feasible and accessible.  
The majority of these advances are 
in situations where the likelihoods 
can not be computed directly and are instead approximated. 
Marginal sampling schemes, like the particle marginal 
Metropolis-Hastings (PMMH) algorithm \citep{Andrieu:uy}, 
alternately sample the structural parameters and latent process parameters 
and accept or reject them with an adjusted Metropolis-Hastings step accounting 
for the approximation of the likelihood needed in sampling the latent space. 
Recent work \citep{BHATTACHARYA2018187} approximates the posterior of the 
structural parameters on a grid of points, reducing the possible sampling 
values to a discrete set. 
SMC has also been adapted to this situation in methods like SMC$^2$ \citep{Chopin:ub}, 
or more aptly named Particle SMC, where one can exploit the known sequential 
structure of otherwise intractable likelihoods.

Our approach takes advantage of factoring the joint posterior density
of $\theta$ and $\bm{\alpha}$ as follows
\begin{align}
	\label{eq:post_break}
    p(\bm{\alpha}, \theta|\bm{y}) = p(\theta|\bm{y})p(\bm{\alpha}|\theta,\bm{y}).
\end{align}
where we omit the dependence on both $\{{\bm{c}}_{it}\}$ and $\bm x_i$.
The first factor in~(\ref{eq:post_break}) is discussed below,
while the second factor, the conditional posterior density 
of the latent process parameters, can be derived exactly and is 
discussed in Section \ref{sub:latent_state_inference}. 

Marginal inference about $\theta$ can be accomplished by 
integrating~(\ref{eq:post_break}) with respect to $\bm{\alpha}$,
yielding the first factor in the expression.
This factor can be expanded using Bayes' Theorem as
\begin{align}
	\label{eq:MDLM_Bayes}
     p(\theta|\bm{y}) = \frac{ p(\bm{y}|\theta) p(\theta)}{ p(\bm{y})}.
\end{align}
The marginal likelihood, $p(\bm{y}|\theta)$, in the numerator is determined from
\begin{align}
\label{eq:marginalize_alpha}
p(\bm{y}|\theta) = \int{p(\bm{y},\bm{\alpha}|\theta)}d\bm{\alpha}.	
\end{align}
As we derive in Section~\ref{sub:marginal_likelihood_for_a_simplified_model},
the marginal distribution in~(\ref{eq:marginalize_alpha})
is multivariate normal (MVN) with a mean and variance that depend only on the 
fixed model parameters $\theta$, the adherence measures $\{{\bm{c}}_{it}\}$ and 
individual covariates $\bm x_i$. 
This marginalization is made possible by normality of the latent state innovations
$\bm\nu_{it}$ and sampling error $\bm\varepsilon_{it}$. 
Inference for $\theta$ can therefore be obtained directly from this marginal
distribution.

\subsection{Marginal Likelihood of the DLM} 
\label{sub:marginal_likelihood_for_a_simplified_model}

We explain our marginalization approach in the case of
a DLM with a single outcome variable ($k=1$)
and for one individual, and assume a single baseline covariate ($p=1$) and
single time-varying covariate ($r=1$) which is assumed to be an adherence indicator.
The parameters in 
$\theta$ are therefore $(\beta, \sigma^2_\varepsilon, \rho, \phi, \sigma^2_\nu, \sigma^2_0)$. 
We assume that both adherence indicators and time-varying outcomes are measured 
over $T$ consecutive days;
the outcomes are
$(y_{1}, y_{2}, ..., y_{T})$ and the adherence indicators are
${\bm{c}} = (c_{1}, c_{2}, ..., c_{{T}})$. 
We can write the vector of outcomes as the sum of a shared non-time-varying component, 
a time-varying component and an error term as
\begin{align}
\label{eq:decomp}
\begin{pmatrix}
    y_{1}\\
    y_{2}\\
    \vdots\\
    y_{T}
\end{pmatrix} = \begin{pmatrix}
    \beta x\\
    \beta x\\
    \vdots\\
    \beta x
\end{pmatrix} + 
\begin{pmatrix}
    \alpha_{1}\\
    \alpha_{2}\\
    \vdots\\
    \alpha_{T}
\end{pmatrix} +  
\begin{pmatrix}
    \varepsilon_{1}\\
    \varepsilon_{2}\\
    \vdots\\
    \varepsilon_{T}
\end{pmatrix} = \beta x\ {\mathds{1}}_T + \bm{\alpha} + \bm{\varepsilon}.
\end{align}
where
we use the ${\mathds{1}}_T$ notation to indicate a column vector 
of length $T$ consisting  of all ones.
The following development is
conditional on $\theta$ and covariates $(x, \bm{c})$ unless noted otherwise.
Thus
the first term on the right hand side of Equation (\ref{eq:decomp}), 
$\beta x\ {\mathds{1}}_T$, 
is treated as a constant vector. 
The third term is distributed as 
\begin{align*}
    \bm{\varepsilon} \sim N_T({\bm{0}}, \sigma_\varepsilon^2 I),
\end{align*} 
where ${\bm{0}}$ is a column vector of zeros, and $I$ is a $T$-dimensional identity matrix. 

Because the initial latent variable $\alpha_1$ is normally distributed, 
and each $\alpha_t$ conditional on the
previous latent variables
is a linear combination of normal random variables, then 
$(\alpha_1, ..., \alpha_T)$ is MVN. 

The AR(1) structure of the latent states admit a recursive mean and variance calculation
\begin{align*}
\mathbb{E}[\alpha_t] & = \rho\mathbb{E}[\alpha_{t-1}] + \phi\ c_t,\\
\mathbb{V}ar[\alpha_t] & =\rho^2\mathbb{V}ar[\alpha_{t-1}] + \sigma_\nu^2.
\end{align*}
This recursion and the initial conditions imply a general formula for 
the mean and variance of $\alpha_t$
\begin{align}
E_t &= \mathbb{E}[\alpha_t] = \phi \sum_{k = 2}^t{\rho^{t-k} c_k} \label{eq:marginal_mean}\\
V_t &= \mathbb{V}ar[\alpha_t] = \sigma_\nu^2 \sum_{k = 2}^t{\rho^{2(t-k)}} + \rho^{2(t-1)}\sigma_0^2 \nonumber
\end{align}
for $t \in \{2,..., T\}$ with initial mean $E_1 = \mathbb{E}[\alpha_1] = 0$ and initial variance $V_1 = \mathbb{V}ar[\alpha_1] = \sigma_0^2$. We collect the mean terms into a vector 
\begin{align}
\label{eq:marginal_mean_vec}
	{\bf{E}} = (E_1, ..., E_T).
\end{align}
The covariance follows a similar recursion, 
\begin{align*}
\mathbb{C}ov(\alpha_t, \alpha_{t-1}) &= \mathbb{C}ov(\rho\alpha_{t-1} + \phi\ c_t + \nu_{t}\ ,\  \alpha_{t-1})\\
&= \rho\ \mathbb{V}ar[\alpha_{t-1}].
\end{align*}
We apply the recursion 
to obtain a general form of the covariance for $t\in \{2,..., T\}$ and $k \in \{0, ..., t-1\}$ 
\[
\mathbb{C}ov(\alpha_t, \alpha_{t-k}) = \rho^k \mathbb{V}ar[\alpha_{t-k}] = \rho^k V_{t-k}.
\]
More compactly, the covariance matrix is
\begin{align}
\label{eq:marginal_cov}
\Sigma = \begin{pmatrix}
    \phantom{-}^{}V_1 & \rho\ V_1 & \rho^2\ V_1 & \dots  & \rho^{T-1}\ V_1 \\
    \rho\ V_1 & \phantom{-}V_2 & \rho\ V_2 & \dots  & \rho^{T-2}\ V_2\\
    \rho^2\ V_1 & \rho\ V_2 & \phantom{-}V_3  & \dots  & \rho^{T-3}\ V_3\\
    \vdots & \vdots & \vdots & \ddots & \vdots \\
    \rho^{T-1}\ V_1 & \rho^{T-2}\ V_2 & \rho^{T-3}\ V_3 & \dots  & \phantom{-}V_T
\end{pmatrix}.
\end{align}
The distribution of the terms in Equation (\ref{eq:decomp}) can be
written explicitly as
\begin{align}
    (y_{1}, y_{2}, ..., y_{T})^T \sim N_T(\beta x\ {\mathds{1}}_T + {\bf{E}}, \Sigma + \sigma_\varepsilon^2 I).
    \label{eq:mvn1}
\end{align} 
The distribution in~(\ref{eq:mvn1}) is proportional to
the marginal likelihood for $T$ sequentially-observed outcomes, 
marginalizing out the latent parameters. 

Let $\mathcal{T} = \{t_1, t_2, ..., t_m\}$ be the {\emph{actual}} observation times, and let 
\begin{align*}
	\bm{E}_\mathcal{T} = (E_{t_1}, E_{t_2}, ..., E_{t_m})^T.
\end{align*}
By properties of the MVN distribution, the outcome vector 
$y_{\bcdot} = (y_{t_1}, y_{t_2}, ..., y_{t_m})^T$ is also MVN with 
mean and covariance respectively
\begin{eqnarray}
\label{eq:marginal_observed_outcomes}
\mathbb{E}[y_{\bcdot}] &=&
\beta x\ {\mathds{1}}_{|\mathcal{T}|} + \bm{E}_\mathcal{T} \nonumber \\
\mathbb{V}ar[y_{\bcdot}] &=&
\Sigma_{\mathcal{T} \times \mathcal{T}} + \sigma_\varepsilon^2 I_{|\mathcal{T}|}.
\end{eqnarray}
The notation $\Sigma_{\mathcal{T} \times \mathcal{T}}$ indicates subsetting 
the covariance matrix $\Sigma$ to the corresponding rows and columns 
designated by $\mathcal{T}$. 
If, for example, $\mathcal{T} = \{1, 7, 9\}$ the first, seventh and ninth 
rows and columns are taken from $\Sigma$ making $\Sigma_{\mathcal{T} \times \mathcal{T}}$ 
a $3\times 3$ matrix. 
One consequence of the marginalization is that if we observe multiple measurements 
of the same outcome on a single day, they will only vary according to 
$\sigma_\varepsilon^2$. 

It is worth noting that even though this marginal distribution does not depend 
on the time-varying parameters,
the autoregressive structure remains in both the 
mean and covariance. 
For example, the centered marginal mean of $y_{t}$ can be shown 
with simple algebraic manipulation
to be  
\begin{align*}
    \left(\mathbb{E}[y_{t}] - \beta x\right)
    &= \rho\left(\mathbb{E}[y_{t-1}] - \beta x\right) + \phi c_t.
\end{align*}
This is the same recursive relationship of the dynamic component of 
our DLM in~(\ref{eq:DLM_dynamic}).  
Even though we marginalize out the time-varying parameters, their sequential structure is retained.

The calculations above were derived for one study participant and a single outcome variable.
These calculations can also be extended to multivariate outcome measurements.
The marginal mean and covariance are both individual- and outcome- dependent, 
as both $\bm\phi_k$ and $\alpha_{itk}$ vary by outcome and individual. 
When referring to multiple study participants and multiple outcome variables,
we can make the dependence
explicit by denoting the quantities in~(\ref{eq:marginal_mean_vec}) 
and~(\ref{eq:marginal_cov}) for outcome $k$ of individual $i$ as ${\bf{E}}_{ik}$ 
and $\Sigma_{ik}$ respectively. 

Because the majority of the structure is contained within outcome (recall $\bm\rho$ 
is a diagonal matrix), we organize the multivariate outcomes as follows. 
We collect the outcome $k$ observed through time $t = 1, ..., T_i$ 
into a $T_i\times 1$ vector $\bm{y}_{i\bcdot k} = (y_{i1k}, y_{i2k}, ..., y_{i{T_i}k})^T$. 
The marginal likelihood of a vector of complete outcome measurements can be written as 
\begin{align}
\label{eq:marginal_multivariate}
\begin{pmatrix}
    \bm{y}_{i\bcdot 1}\\
    \bm{y}_{i\bcdot 2}\\
    \vdots\\
    \bm{y}_{i\bcdot k}\\
\end{pmatrix}\ |\ \theta
\sim N_{kT} \left(
\bm{E}_i +  \bm\beta \bm{x}_i  \otimes {\mathds{1}}_{T_i}
, \;
\Sigma_i + \Sigma_\varepsilon \otimes  I_{T_i}
\right)
\end{align}
where $\otimes$ is the Kronecker product, and where
\begin{align}
\label{eq:marginal_multivariate_def}
	{\bm{E}}_i = \begin{pmatrix}
    {\bf{E}}_{i1}\\
    {\bf{E}}_{i2}\\
    \vdots\\
    {\bf{E}}_{ik}\\
\end{pmatrix}\quad \text{and}\quad \Sigma_i = \begin{pmatrix}
    \Sigma_{i1} &\bm{0} & \cdots & \bm{0} \\
    \bm{0} & \Sigma_{i2} & \cdots & \bm{0} \\
    \vdots & \vdots & \ddots & \vdots \\    
    \bm{0} &\bm{0} & \cdots & \Sigma_{ik} \\
\end{pmatrix}.
\end{align}
As in~(\ref{eq:marginal_observed_outcomes}), 
we can derive the marginal distribution of the outcome measures 
$\bm{y}_{i\bcdot\bcdot}|\theta$ by subsetting the appropriate rows 
and columns of the quantities in~(\ref{eq:marginal_multivariate_def}).

If we assume that outcomes between
study participants are 
independent conditional on $\theta$, 
the marginal distribution of outcomes unconditional on the
time-varying parameters is given by
\begin{align}
	p\left(\bm{y}_{1\bcdot\bcdot}, ..., \bm{y}_{n\bcdot\bcdot}|\theta, {\bm{x}},{\bm{c}}\right) =  \prod_{i = 1}^n p\left(\bm{y}_{i\bcdot\bcdot}|\theta, x_i,{\bm{c}}_{i\bcdot\bcdot}\right).
	\label{eq:multiple_ys}
\end{align}
This simple marginalization drastically reduces the number of parameters in 
the model and simplifies computation. 
We can conduct inference on our non-dynamic parameters 
in a Bayesian setting by introducing a prior distribution for $\theta$
and using the marginal likelihood of $\theta$ which is proportional
to~(\ref{eq:multiple_ys}).
The posterior density for the non-dynamic parameters is given by
\begin{align}
\label{eq:structural_posterior}
	p\left(\theta|\left\{\bm{y}_{i\bcdot\bcdot}\right\}_{i = 1}^n, {\bm{x}},{\bm{c}}\right) \propto p(\theta) \prod_{i = 1}^n p\left(\bm{y}_{i\bcdot\bcdot}|\theta, x_i,{\bm{c}}_i\right).
\end{align}
Inference can be performed in a straightforward and efficient way 
via MCMC, or 
using Hamiltonian Monte Carlo \citep{homan_no-u-turn_2014} sampling
to obtain draws from the posterior distribution, as implemented in
STAN \citep{carpenter_stan:_2017}.
Alternatively, posterior sampling could be obtained 
using Sequential Monte-carlo (SMC) in software packages like Libbi \citep{2013arXiv1306.3277M}. 
SMC works with the sequential likelihoods $p(y_{it_j}|y_{it_1}, ..., y_{i t_{j-1}}, \theta)$, which are easily available in our framework because they are conditional distributions of the Multivariate Normal distribution in~(\ref{eq:marginal_multivariate}).

\subsection{Inference for the Latent Process} 
\label{sub:latent_state_inference}

Posterior inference for $\bm{\alpha}$ can be determined once 
$p(\theta|\bm{y})$ 
has been obtained 
by exploiting the factorization of the posterior density in~(\ref{eq:post_break}).
To do so, we note that
the joint distribution of $\bm{\alpha}$ and $\bm{y}$ conditional on $\theta$
is MVN. 
For our development,
we consider the case in which outcomes are observed for $T$ consecutive days.
With a $k$-dimensional outcome variable,
$\bm{\alpha}$ contains $kT$ elements and $\bm{y}$ contains $kT$ (scalar) outcomes.
Using the notation in~(\ref{eq:marginal_multivariate_def}), 
the joint distribution of outcomes and time-varying parameters is
\begin{align}
\label{eq:latent_post_conditional}
\left.
\begin{pmatrix}
    \bm{\alpha}_{i\bcdot 1}\\
    \vdots\\
    \bm{\alpha}_{i\bcdot k}\\
    \bm{y}_{i\bcdot 1}\\
    \vdots\\
    \bm{y}_{i\bcdot k}\\
\end{pmatrix}\ \right|\ \theta
\sim N_{2kT} \left(
\begin{pmatrix}
    {\bf{E}}_i\\
    \\
    {\bf{E}}_i + \bm\beta \bm{x}_i  \otimes {\mathds{1}}_{T_i}\\
\end{pmatrix}
, 
\begin{pmatrix}
    \Sigma_i & \Sigma_i \\
    \\
    \Sigma_i & \Sigma_i+ \Sigma_\varepsilon \otimes  I_{T_i} \\
\end{pmatrix} 
\right).
\end{align}
The quantities on the diagonal of the covariance matrix 
in~(\ref{eq:latent_post_conditional}) were determined in 
Section~\ref{sub:marginal_likelihood_for_a_simplified_model}. 
The covariance of $\bm{y}_{i\bcdot k}$ and $\bm{\alpha}_{i\bcdot \ell}$ 
(for $k \ne \ell$),
conditional on $\theta$, can be shown to be $\bm{0}$ as follows
(with conditioning on $\theta$ suppressed).
\begin{align*}
    \mathbb{C}ov(\bm{\alpha}_{i\bcdot l}, \bm{y}_{i\bcdot k}) &= 
     \mathbb{E}\left[\mathbb{C}ov\left(\bm{\alpha}_{i\bcdot l}, \bm{y}_{i\bcdot k}| \bm{\alpha}_{i\bcdot k} \right)\right] 
     + \mathbb{C}ov\left(\mathbb{E}[\bm{\alpha}_{i\bcdot l}| \bm{\alpha}_{i\bcdot k}], \mathbb{E}[\bm{y}_{i\bcdot k} | \bm{\alpha}_{i\bcdot k}]\right) \\
    &= \mathbb{E}\left[\mathbb{C}ov\left(\bm{\alpha}_{i\bcdot l}, \beta x_i + \bm{\alpha}_{i\bcdot k} + \bm\varepsilon_{i\bcdot k} | \bm{\alpha}_{i\bcdot k} \right)\right] 
    + \mathbb{C}ov\left(\bm{\alpha}_{i\bcdot l}, \beta x_i + \bm{\alpha}_{i\bcdot k}\right) \\
    &= \mathbb{E}\left[\mathbb{C}ov\left(\bm{\alpha}_{i\bcdot l}, \bm\varepsilon_{i\bcdot k} \right)\right] 
    + \mathbb{C}ov\left(\bm{\alpha}_{i\bcdot l}, \bm{\alpha}_{i\bcdot k}\right) \\ &= \bm{0}.
\end{align*}
We can subset the mean vector and covariance matrix of~(\ref{eq:latent_post_conditional}) 
as in~(\ref{eq:marginal_observed_outcomes}) to account for the sporadically observed outcomes, 
but we continue to assume that the response values are observed at each time period.
The conditional posterior distribution of the latent parameters is determined from
the joint distribution in~(\ref{eq:latent_post_conditional})
by conditioning on $\bm{y}$ as 
\begin{align}
\label{eq:latent_posterior_2D}
\left.
\begin{pmatrix}
    \bm{\alpha}_{i\bcdot 1}\\
    \vdots\\
    \bm{\alpha}_{i\bcdot k}\\
\end{pmatrix} \right| \bm{y}_{i\bcdot 1}, ..., \bm{y}_{i\bcdot K}, \theta \sim N_{kT}\left( \tilde{\alpha}_i, 
\tilde\Sigma_i\right)
\end{align}
where 
\begin{align*}
    \tilde\alpha_i = \left({\bf{E}}_i\right) + \Sigma_i \left(\Sigma_i+ \Sigma_\varepsilon \otimes  I_{T_i}\right)^{-1}
\left[
\begin{pmatrix}\bm{y}_{i\bcdot 1}\\
    \vdots\\
    \bm{y}_{i\bcdot k}\\
\end{pmatrix} - \left(
{\bf{E}}_i + \bm\beta \bm{x}_i  \otimes {\mathds{1}}_{T_i}\right)
\right]
\end{align*}
and 
\begin{align*}
    \tilde\Sigma_i = \Sigma_i - \Sigma_i  \left(\Sigma_i+ \Sigma_\varepsilon \otimes  I_{T_i}\right)^{-1}\Sigma_i.
\end{align*}
We can use the joint distribution specified in~(\ref{eq:latent_posterior_2D}) 
along with the posterior distribution on the structural parameters 
shown in~(\ref{eq:structural_posterior}) to sample from the latent process parameters 
by first simulating 
$\theta^* \sim p\left(\theta|\bm{y}_{1\bcdot\bcdot}, ..., \bm{y}_{n\bcdot\bcdot}, {\bm{x}},{\bm{c}}\right)$ 
and then sampling from $p(\bm{\alpha}_{i\bcdot 1}, ..., \bm{\alpha}_{i\bcdot 1}| \bm{y}_{1\bcdot\bcdot}, ..., \bm{y}_{n\bcdot\bcdot}, \theta^*)$. 

\section{Application to BP Study} 
\label{sec:application_to_bp_study}

We return to an application of our framework to modeling time-varying BP 
as a function of time-varying adherence to anti-hypertensive medication 
and baseline comorbidities and demographics described in Section \ref{sec:data}. 
We begin with a discussion of models often used in this task that 
incorporate adherence as non-dynamic information. 
We also address missing adherence measures for a fraction of 
patients. 

\subsection{Non-dynamic models incorporating adherence}
\label{subsec:nondynamic_adhere}

Adherence to medication is typically incorporated into outcome models 
in one of two ways. 
The average adherence for the study period is either used directly or 
is dichotomized into two groups, those below a certain threshold and those above, 
indicating ``poor'' versus ``good'' adherence \citep{rose_effects_2011, doi:10.1001/jama.296.21.joc60162, doi:10.1001/archinte.164.7.722}. 
Either of these approaches includes adherence as a non-dynamic covariate in the model.
Repeated outcome measures are accompanied by patient-specific random effects.

We present these
alternative models in the context of BP outcomes, a bivariate measure.
We label the outcomes $1$ for Systolic BP and $2$ for Diastolic BP. 
These models take the form
\begin{align}
	\label{eq:alternative_likelihood}
	\begin{pmatrix}y_{it1} \\ y_{it2}\end{pmatrix} = 
	\begin{pmatrix}x_i\beta_1 \\ x_i\beta_2\end{pmatrix} + 
	\begin{pmatrix}\delta_{i1} \\ \delta_{i2}\end{pmatrix} + 
	\begin{pmatrix}\bar{c}_i \gamma_1 \\ \bar{c}_i \gamma_2  \end{pmatrix}  + \begin{pmatrix} \varepsilon_{it1}\\ \varepsilon_{it2} \end{pmatrix}
\end{align}
with two different adherence measures, $\bar{c}_i$.  
We first consider average adherence, $\bar{c}_i = T_i^{-1} \sum_{t=1}^{T_i} c_{it}$ 
over $T_i$ days for patient~$i$.
In this case we would interpret the adherence effect parameters 
$(\gamma_1, \gamma_2)$ as the differences in BP of being fully adherent 
relative to being fully non-adherent, controlling for the baseline covariates.
We also consider a dichotomized summary 
$\bar{c}_i = \mathds{1}\left\{\left(T_i^{-1} \sum_{t=1}^{T_i} c_{it}\right) > 2p-1\right\}$, 
an indicator of overall adherence. 
In this case $(\gamma_1, \gamma_2)$ would be interpreted as the differences in 
BP for those with ``good'' ($\bar{c}_i = 1$) versus ``poor'' ($\bar{c}_i = 0$) adherence. 
Several values of $p$ were considered to assess the sensitivity to this choice, 
but we present results for $p=0.8$ which is a
conventional choice \citep{doi:10.1001/archinte.164.7.722}.  
The model in Equation (\ref{eq:alternative_likelihood}) includes
patient-specific random effects $\delta_{i1} \sim N(0,\sigma_{\delta 1}^2)$ and 
$\delta_{i2} \sim N(0,\sigma_{\delta 2}^2)$. 

We compare the fit of the above two non-dynamic models to our dynamic model
framework.
The non-dynamic approach has a potential advantage of being more robust to 
detailed model misspecification relative to our dynamic model, particularly
with the choice of the specific model for the evolution of the health measures
as a function of daily adherence.
However, incorporating average adherence may mask important time-varying effects
of detailed adherence.
We explore this tradeoff in Section~\ref{subsec:data_analyses}.

\subsection{Missing adherence indicators}
\label{subsec:missing_adherence}

Not all of the patients in our cohort have completely observed adherence indicators.
Among the $503$ patients in our analyses,
$70$ have at least one day of missing adherence.  
Adherence could be missing due to MEMS cap malfunctions, hospital inpatient stays 
in which the MEMS containers were not used,
or other causes. 
The majority ($50$) of the $70$ patients had only
one or two missing adherence values.  
In the most extreme case one patient was missing $48$ out of $102$ adherence measures.  

For patient $i$, let $\bm{c}_{i}^{obs} = \{c_{it}\}_{t\in\mathcal{T}_{obs}}$ be the set of observed 
adherence values and $\bm{c}_{i}^{mis} = \{c_{it}\}_{t\in\mathcal{T}_{mis}}$ be the set of missing 
adherence values. 
Letting $\eta_i$ represent the parameters describing a potential adherence 
model for patient $i$, the posterior density in~(\ref{eq:structural_posterior}) 
conditional on the observed adherence data only
is
\begin{align}
\label{eq:adherence_impute}
	p\big(\theta|{\bm{y}}_{i\bcdot\bcdot}, x_i, \bm{c}_{i}^{obs} \big)
	&= \int{p\big(\theta, \bm{c}_{i}^{mis}, \eta_i|{\bm{y}}_{i\bcdot\bcdot}, x_i, \bm{c}_{i}^{obs} \big) }\  d\bm{c}_{i}^{mis}\ d\eta_i  \nonumber\\
	&= \int{p\big(\theta|{\bm{y}}_{i\bcdot\bcdot}, x_i, \bm{c}_{i}^{obs}, \bm{c}_{i}^{mis}, \eta_i \big)
	p\big(\bm{c}_{i}^{mis}|{\bm{y}}_{i\bcdot\bcdot}, x_i, \bm{c}_{i}^{obs}, \eta_i \big) p(\eta_i|{\bm{y}}_{i\bcdot\bcdot}, x_i, \bm{c}_{i}^{obs}) }\  d\bm{c}_{i}^{mis}\ d\eta_i  \nonumber\\
    &\stackrel{(a)}{=}  \int{p\big(\theta|{\bm{y}}_{i\bcdot\bcdot}, x_i, \bm{c}_{i}^{obs}, \bm{c}_{i}^{mis}, \eta_i \big)
	p\big(\bm{c}_{i}^{mis}| \eta_i \big) p(\eta_i| \bm{c}_{i}^{obs}) }\  d\bm{c}_{i}^{mis}\ d\eta_i .
\end{align}
The last line in~(\ref{eq:adherence_impute}) labeled $(a)$ involves a set of modeling choices. 
First, we assume that the adherence model parameters do not depend on other covariates
or the outcomes, and this is reflected 
by assuming $p(\eta_i|{\bm{y}}_{i\bcdot\bcdot}, x_i, \bm{c}_{i}^{obs}) = 
p(\eta_i| \bm{c}_{i}^{obs})$. 
Other approaches, such as those described by \citet{Naranjo:2013io}, 
provide a framework 
for modeling missing time-varying covariates in DLMs with complex models. 
However, our assumption is
conservative but likely more robust to model misspecification.
Second, $p\big(\bm{c}_{i}^{mis}|{\bm{y}}_{i\bcdot\bcdot}, x_i, \bm{c}_{i}^{obs}, \eta_i \big) = p\big(\bm{c}_{i}^{mis}| \eta_i \big)$, 
implies that the missing adherence values can be simulated directly from our adherence model, 
represented by $\eta_i$, without regard to individual-level characteristics. 
Again, this is a conservative choice because adding information to the imputation model could
help improve predictions if properly modeled.

We simulate the missing adherence measures from Beta-Bernoulli distributions that
depend only on adherence measures for each patient separately.
Letting $\eta_i$ represent patient $i$'s overall adherence, 
the unobserved adherence values are drawn from a 
Bernoulli distribution
$$\{c_{it}^*\}_{t\in\mathcal{T}}|\eta_i \stackrel{i.i.d.}{\sim} 
\text{Bernoulli}(\eta_i)$$
where the $c_{it} = 2c_{it}^* - 1$ take on
values $\{-1, 1\}$. 
Under a uniform prior distribution on patient $i$'s adherence rate,
the posterior distribution for $\eta_i$ is given by
$$\eta_i|\{c_{it}\}_{t\in\mathcal{T}_{obs}} \sim \mbox{Beta}(n_{i,1}+1,n_{i,-1}+1)$$ 
when they were observed to be adherent 
$n_{i,1}$ days and non-adherent $n_{i,-1}$ days. 
 
We repeated the simulation process $20$ times and combined the 
posterior samples for a proper Bayesian multiple-imputation analysis 
\citep{little_statistical_2002} for our non-dynamic models.
For our fully Bayesian model, the simulated adherence values were incorporated
into our posterior simulation analyses.
We found that in both cases
inference of our non-dynamic parameters $\theta$ was not sensitive 
to the missing adherence values.

\subsection{Analysis of BP measures}
\label{subsec:data_analyses}

Table \ref{tab:BP_readings} displays the number of BP measurements 
among our $503$ patients during the study period.
While a majority ($69.8\%$) had either one or two BP readings, a substantial 
number of patients had $3$ or more. 
The number of BP readings totalled $1152$, averaging $2.29$ readings per patient. 
\begin{table}[ht]
\centering
\begin{tabular}{r||r|r|r|r|r|r|r|r|}
 Number of BP Readings& $1$ & $2$ & $3$ & $4$ & $5$ & $6$ & $7$ & $8+$ \\ 
  \hline
Number of Patients & $226$ & $125$ & $72$ & $28$ & $16$ & $12$ & $15$ & $9$ \\ 
  Percent of Patients & $44.90$ & $24.90$ & $14.30$ & $5.60$ & $3.20$ & $2.40$ & $3.00$ & $1.79$\\ 
\end{tabular}
\caption{
Number of BP readings during the study period.
\label{tab:BP_readings}
}
\end{table}

For the period of the study, the proportion of days that patients took their 
medication varied widely. 
The proportions ranged from $8.9\%$ to $100\%$, with a median of $95.1\%$ and a
mean of $88.9\%$.
These adherence rates on average were high, with many patients being fully adherent
throughout the study period.
Only $19.1\%$ of patients had below $80\%$ adherence. 
The high degree of adherence is consistent with recruiting patients for the study
who were continual users of anti-hypertensive medication.
On average, adherence was recorded over
$98$ days per patient (minimum and maximum of $21$ and $395$ days, 
respectively, with an inter-quartile range of $14$), 
with $75\%$ of 
patients followed between $84$ and $112$ days. 

Baseline summaries of the non-dynamic covariates appear 
in Table \ref{tab:baseline}. 
\begin{table}[ht!]
\centering
\begin{tabular}{r|r}
& Mean (Std Dev) \\ 
  \hline
Age (y)& $60.3 (11.1)$\\
DBP at Enrollment& $79.7 (11.5)$\\
SBP at Enrollment& $133.5 (19.6)$\\
\hline
\hline
 & Percent\\
\hline
Female& 67.5\\
African-American& 54.9\\
Income below $\$20,000$& 44.3\\
Obese& 60.6\\
Cerebral Vascular Disease& 5.6\\
Congestive Heart Failure& 3.8\\
Renal Insufficiency& 6.8\\
Coronary Artery Disease& 14.9\\
Diabetes& 36.2\\
Hyperlypidemia& 57.1\\
Peripheral Vascular Disease& 6.2\\
\end{tabular}
\caption{Baseline socio-demographic and health characteristics of the patients in the study cohort.}
\label{tab:baseline}
\end{table}
A majority of the cohort consisted of women, more
patients of black race (African or Caribbean descent) than white, 
and a large fraction of low-income patients. 
The cohort also consisted of mostly obese patients, 
and had a moderately high comorbidity burden. 
Based on 503 patients with BP readings within 14 days of enrollment, 
the cohort on average had relatively well-controlled hypertension 
at baseline, as all patients were prescribed 
anti-hypertensive medication (though adherent to varying extents). 
The cohort consisted of  $26.2\%$ having 
DBP$>$80 mm Hg and SBP$>$130 mm Hg), 
and $7.8\%$ having
DBP$>$90 mm Hg and SBP$>$140 mm Hg).

A prior distribution was assumed for the alternative models 
specified in~(\ref{eq:alternative_likelihood}). 
For $k=1,2$, indicating systolic and diastolic BP measures
separately, we assumed the following.
\begin{align*}
\rho_{k} & \sim U(-1,1)\\
\phi_{k} & \sim N(0,25)\\
\gamma_k &\sim N(0, 25)\\
\sigma_{\varepsilon k} &\sim U(0, 30), \rho_\varepsilon  \sim U(-1,1)\\
\sigma_{\nu k} &\sim U(0, 10), \sigma_{0 k} \sim U(0, 30)\\
\beta_{11}&\sim N(120, 400), \beta_{12}\sim N(80, 400)\\
\beta_{jk}&\sim N(0, 400), j = 2, ..., p\\
\sigma_{U k} &\sim U(0, 30)
\end{align*}
The prior components were selected to be vague but proper.
The intercepts $\beta_{1k}$ had distributions centered near 
the typical systolic and diastolic BPs, but 
had variances that were sufficiently large to 
acknowledge the uncertainty in the effects.
We assumed uniform prior components with compact support for
the standard deviation parameters, as recommended by
\citet{gelman2006prior}.
The correlation and autocorrelation parameters were 
assumed to have uniform prior components as in the dynamic model
parametrization. 
Convergence of the MCMC simulated values was inspected with trace plots 
of multiple chains, and using the Gelman-Rubin convergence 
statistic \citep{gelman1992}. 

Table \ref{tab:covariate_effects} presents 
posterior means and 90\% central posterior intervals
for the non-dynamic covariate effects 
using the DLM. 
\begin{table}[ht!]
\centering
\begin{tabular}{|l|rrrr|}
\hline
Variable & Systolic & ($90\%$ CI) & Diastolic & ($90\%$ CI)\\ \hline
Intercept & $132.97$ & $(129.46, 136.51)^{\star\dagger\ddagger}$ & $85.68$ & $(83.65, 87.72)^{\star\dagger\ddagger}$\\
Sex (male) & $-1.1$ & $(-3.59, 1.38)$ & $0.53$ & $(-0.88, 1.94)$\\
Age (group 1) & $-0.01$ & $(-3.3, 3.27)$ & $-1.5$ & $(-3.4, 0.4)$\\
Age (group 2) & $1.87$ & $(-1.45, 5.19)$ & $-3.54$ & $(-5.45, -1.63)^{\star\dagger\ddagger}$\\
Age (group 3) & $5.22$ & $(1.57, 8.89)^{\star\dagger\ddagger}$ & $-6.73$ & $(-8.83, -4.62)^{\star\dagger\ddagger}$\\
White & $-3.31$ & $(-5.61, -1.01)^{\star\dagger\ddagger}$ & $-1.58$ & $(-2.88, -0.26)^{\star}$\\
Obese & $3.07$ & $(0.76, 5.38)^{\star\dagger\ddagger}$ & $1.68$ & $(0.35, 3.01)^{\star\dagger\ddagger}$\\
Nicotine dependence & $-0.9$ & $(-5.22, 3.42)$ & $1.3$ & $(-1.17, 3.76)$\\
Hyperlipidemia & $-1.44$ & $(-3.71, 0.83)$ & $-1.34$ & $(-2.64, -0.04)^{\star}$\\
Diabetes & $1.44$ & $(-0.96, 3.85)$ & $-2.98$ & $(-4.34, -1.61)^{\star\dagger\ddagger}$\\
Peripheral vascular disease & $-1.28$ & $(-5.75, 3.17)$ & $-2.6$ & $(-5.13, -0.06)^{\star\dagger}$\\
Renal insufficiency & $-0.67$ & $(-4.88, 3.52)$ & $-2.73$ & $(-5.14, -0.33)^{\star\dagger\ddagger}$\\
Benign prostatic hypertrophy & $2.82$ & $(-3.66, 9.3)$ & $-1.59$ & $(-5.34, 2.15)$\\
Coronary artery disease & $-1.66$ & $(-4.77, 1.45)$ & $-2.76$ & $(-4.54, -0.98)^{\star\dagger\ddagger}$\\
Congestive heart failure & $-0.78$ & $(-6.05, 4.46)$ & $0.59$ & $(-2.44, 3.59)$\\
Cerebral vascular disease & $2.45$ & $(-2.19, 7.06)$ & $-0.17$ & $(-2.82, 2.48)$\\
\hline
\end{tabular}
\caption{Summaries of covariate effects 
for the bivariate dynamic linear model.
$^\star$Effect with 90\% posterior interval not containing
0 in DLM, 
$^\dagger$Significant effect at the 0.1 level
in average adherence model, 
$^\ddagger$Significant effect at the 0.1 level
in dichotomized adherence model.}
\label{tab:covariate_effects}
\end{table}
The point estimates and intervals are reported for the DLM only;
the effects in the alternative models that were significant at
the $0.1$ level
are indicated with a $\dagger$ (for the average adherence model)
or a $\ddagger$ (for the dichotomized adherence model).
Effects with 90\% central posterior intervals not containing 0
marked with an asterisk ($\star$).
Based on the model fits,
the estimated covariate effects tend to be similar 
across all models with the point estimates tending to agree in 
magnitude and sign. Even though the DLM covariate effects tended 
to have narrower intervals on average ($3\%$ reduction), 
the significance of the findings tended to agree as well.
In particular, the effect of race (white versus non-white)
was significantly negative,
indicating that whites tended to have lower blood pressure 
controlling for all other variables and time-varying adherence. 
Patients who were obese at the beginning of the study tended 
to have significantly higher DBP and SBP.
These findings are consistent with the results of previous studies 
\citep{kressin_understanding_2010,rose_effects_2011}
in their significance and direction of the effects.
Both of these, except for the effect of being white on mean 
diastolic BP, agreed with the alternative models 
in terms of significance and directionality of the effect.

Table \ref{tab:measurement_error} reports inferences for
the standard error and correlations of 
$(\bm\varepsilon_{it})$ for all three models. 
\begin{table}[ht!]
\centering
\begin{tabular}{|l|llll|}
\hline
\multicolumn{5}{|c|}{Dynamic linear model} \\ \hline
Variable & Systolic & ($90\%$ CI) & Diastolic & ($90\%$ CI)\\ \hline
Standard Errors: & $13.37$ & $(12.46, 14.25)$ & $8.11$ & $(7.68, 8.54)$ \\
Correlation: & $0.62$ & $(0.57, 0.67)$ &  & \\
\hline\hline
\multicolumn{5}{|c|}{Average Adherence Model} \\ \hline
Variable & Systolic & ($90\%$ CI) & Diastolic & ($90\%$ CI)\\ \hline
Standard Errors: & 14.53 & (13.87, 15.21) & 8.34 & (7.97, 8.72)\\
Correlation: & 0.58 & (0.54, 0.62) &  & \\
\hline\hline
\multicolumn{5}{|c|}{Dichotomized Adherence Model} \\ \hline
Variable & Systolic & ($90\%$ CI) & Diastolic & ($90\%$ CI)\\ \hline
Standard Errors: & 14.5 & (13.86, 15.18) & 8.34 & (7.96, 8.72)\\
Correlation: & 0.58 & (0.54, 0.62) &  &  \\
\hline\end{tabular}
\caption{Measurement error estimates for three models.}
\label{tab:measurement_error}
\end{table}
Comparing the sampling standard deviation estimates across models 
provides an indication of the gains in modeling the adherence effects as time-varying. 
The standard error estimates for the alternative adherence models tend to be slightly larger 
than those given by the DLM, which is consistent with previous work \citep{rose_effects_2011}. 
The time-varying adherence explicitly captured in the DLM may account for the extra variation
in the outcomes of the alternative adherence models through a reduction in the estimated
measurement error variance.

Table \ref{tab:adherence_effects} contains the adherence effects for our models. 
\begin{table}[ht!]
\begin{tabular}{|l|llll|}
\hline
\multicolumn{5}{|c|}{State Space Model} \\ \hline
Variable & Systolic & ($90\%$ CI) & Diastolic & ($90\%$ CI)\\ \hline
Adherence effect: & -0.48 & (-0.84, -0.2)$^\star$  & -0.24 & (-0.43, -0.09)$^\star$  \\
Asymptotic Adherence effect: & -3.87 & (-5.98, -1.83)$^\star$  & -3.15 & (-4.38, -1.94)$^\star$  \\
\hline\hline
\multicolumn{5}{|c|}{Alternative Adherence Model: Average Adherence} \\ \hline
Variable & Systolic & ($90\%$ CI) & Diastolic & ($90\%$ CI)\\ \hline
Adherence effect: & -9.24 & (-16.03, -2.22)$^\star$  & -9.46 & (-13.55, -5.52)$^\star$ \\
\hline\hline
\multicolumn{5}{|c|}{Alternative Adherence Model: Dichotomized Adherence ($p = 0.8$)} \\ \hline
Variable & Systolic & ($90\%$ CI) & Diastolic & ($90\%$ CI)\\ \hline
Adherence effect: &-5.49 & (-8.26, -2.75)$^\star$  & -3.78 & (-5.27, -2.2)$^\star$ \\
\hline\end{tabular}
\centering
\caption{Adherence effects for the models considered
}
\label{tab:adherence_effects}
\end{table}
The adherence effects across the three models
are not directly comparable, given the different approaches to incorporate adherence.
The average adherence model indicates that the difference between those who were fully adherent 
and those who were fully non-adherent, controlling for other covariates, 
is about $-9.2$ and $-9.5$ for systolic and diastolic, respectively.  
This implies, for example,
that a 10\% additive increase in adherence corresponds to a $0.92$ and $0.95$ 
reduction in systolic and diastolic blood pressure, respectively. 
However, this effect size assumes that the relationship between average adherence and 
blood pressure is linear and holds throughout the entire range of adherence. 
Given the limited range of average adherence observed in the data ($90\%$ 
of patients have average adherence above $70.5\%$),
interpreting this effect beyond this range is not recommended because it 
involves extrapolating beyond the data. 
The dichotomized adherence model shows similar results. 
In particular this approach involves comparing those with relatively good overall adherence 
(above 80\% adherent) to everyone else. 
Based on the dichotomized adherence model,
the benefit of being in the former group is indicated by a lower blood 
pressure of $-5.49$ and $-3.78$ on average for systolic and diastolic
blood pressure.

The DLM gives similar results. 
The effect of taking medication can be inferred on a daily basis. 
The results of the model fit suggest a
small but significant reduction of blood-pressure from taking the medication on a daily basis, 
$-0.24$ mm Hg and $-0.12$ mm Hg on average for systolic and diastolic BP, respectively.  
These estimates imply that, accounting for the correlation estimate,
a patient who is adherent over consecutive days would experience a long-term reduction in systolic BP 
by $-3.9$ mm Hg, and a long-term reduction in diastolic BP by $-3.15$ mm Hg.  
The magnitudes of an increase in long-term BP for continued non-adherence is the 
same but in the opposite direction. 
Overall, the adherence effects tend to agree in terms of the significance and 
direction for the different models. 
However, the DLM provides a clearer interpretation of these parameters that is consistent 
with the time-varying nature of the data.

\begin{figure}
    \begin{center}
        \includegraphics[scale=0.6]{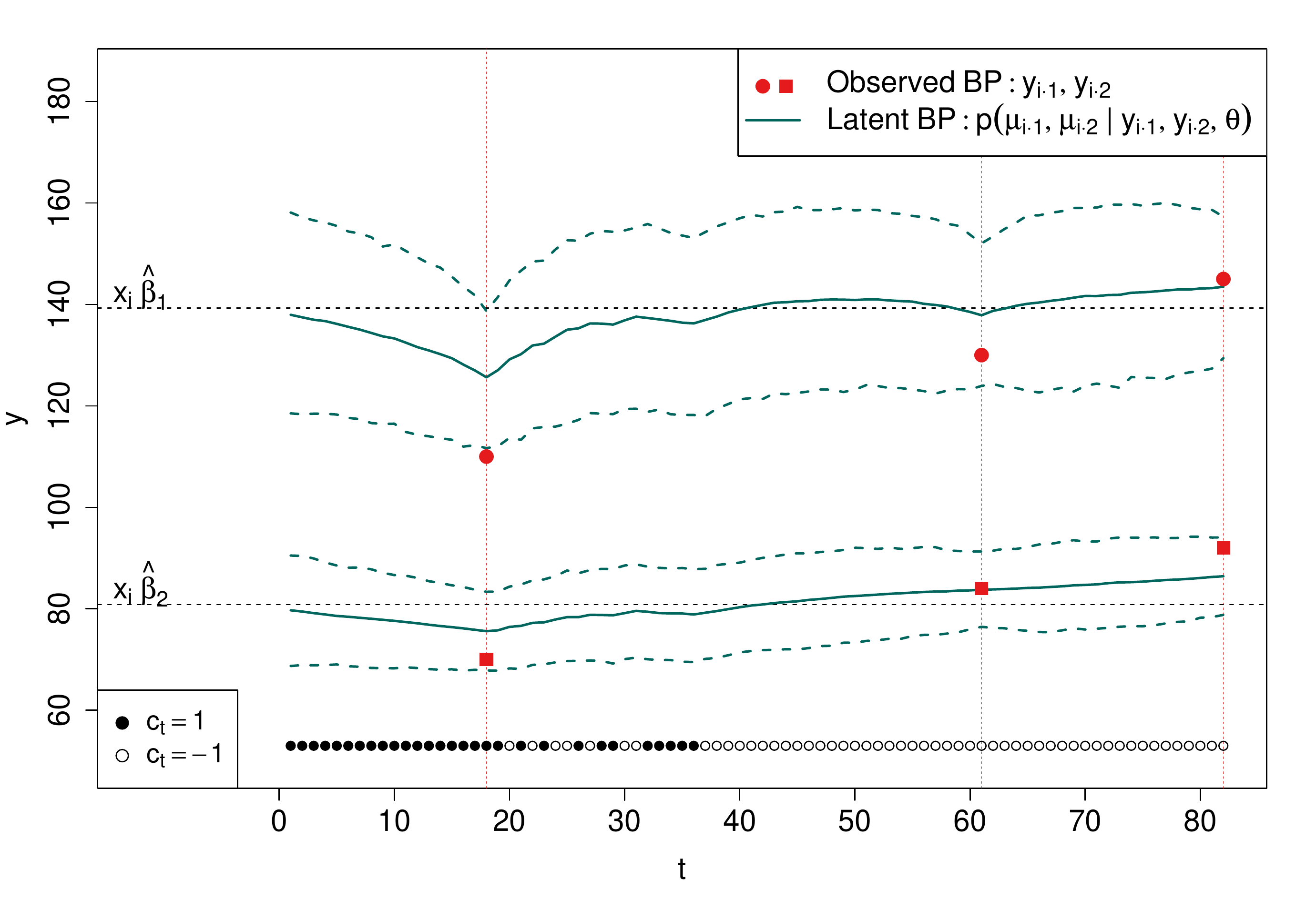}
    \end{center}
    \caption{90\% central posterior intervals for the mean blood pressure $\{\mu_{it\bcdot}\}$ 
    for one study subject.  
    The solid lines are the estimated mean DBP and SBP, and the dashed lines are the 
    point-wise posterior intervals.  Horizontal dotted lines are drawn at the baseline-estimated
    mean DBP and SBP.}
    \label{fig:ssm_post} 
\end{figure}

Another benefit of the DLM is our ability to infer the latent BP for unobserved days 
using the procedure discussed in Section~\ref{sub:latent_state_inference}. 
Figure \ref{fig:ssm_post} contains an example of posterior draws of the 
mean process
$\bm\mu_{it\bcdot} = \bm\beta \bm{x}_i + \bm\alpha_{it\bcdot}$ for the patient 
presented in Figure~\ref{fig:data_example}. 
The solid line indicates the posterior mean of the latent 
process
while the dashed lines 
indicate a $90\%$ credible envelope across time. 
That is, for each time $t$, a $90\%$ credible interval of 
$\bm{\alpha}_{it1}, \bm{\alpha}_{it2}|\left\{\bm{y}_{i\bcdot\bcdot}\right\}_{i = 1}^{n}$ 
is shown. 
The estimated mean DBP and SBP processes vary gradually over time, with shifts in direction
indicated by medication adherence variation.

\section{Discussion} 
\label{sec:discussion}
In this article we propose a multivariate DLM that 
can be used for modeling time-varying outcome measures as a function of detailed
medication adherence or other time-varying covariates. While DLMs are of common use, 
our particular setting benefits from a two-stage computational approach made possible 
by factorizing the posterior density into the dynamic and non-dynamic parameters. 
Typical analyses in this setting ignore the time-varying structure of medication
adherence and instead examine this relationship between adherence and outcomes via 
correlations with non-dynamic adherence measures (e.g., time-averaged adherence). 
Our framework explicitly provides a 
measure of daily impact of medication-taking and meaningful bounds for mean 
BP while controlling for baseline covariates.

Our modeling approach is sufficiently flexible to permit a wide range of
assumptions distinct from those we included in our hypertension application.
For example, we assumed an AR(1) modeling structure for the mean outcome
process, and this assumption can be justified based on the
pharmacokinetic properties of the medications.
In other settings, alternative mean processes can be considered, including
higher order AR processes, growth curves, and so on.
Our baseline covariates were modeled linearly, but 
our framework permits non-linear inclusion of covariate information for
both baseline non-dynamic covariates, as well as the dynamic predictors in the
mean process component of our model.
A crucial assumption of our framework is that the outcome distribution is
multivariate normal, as this assumption allows for the marginalization strategy
that leads to an efficient computational procedure.
However, with some non-normal outcome distributions, various strategies
can be employed that can potentially take advantage of the marginalization idea.
For example, non-normal outcome densities can be approximated by normal distributions
after which the marginalization can be performed; then a Metropolis-Hastings 
algorithm would be incorporated into the posterior sampling procedure that would
account for the non-normality of the original sampling distribution.
Such a procedure is likely to be far more efficient than sampling the
dynamic parameters directly as part of model fitting.

The proposed modeling framework also permits extensions that are straightforward 
to incorporate. For example, multivariate outcomes that are recorded at 
staggered intervals can be accommodated by marginalizing the distribution 
in~(\ref{eq:marginal_multivariate}) appropriately.
We may also believe that the effect of adherence to medication varies from person to person. 
We can then include patient-specific 
adherence effects with a hierarchical prior on 
these parameters to share information across individuals. 
Our analyses did not acknowledge differences among anti-hypertensive 
medications, but our framework
permits distinguishing differential medication effects in a natural manner.
The effects of different medications, perhaps grouped by 
relevant characteristics such as whether the medication is short-acting
or long-lasting, or by medication type (e.g., for hypertension, diuretics, 
ACE inhibitors, etc.), can be included as separate time-varying effects
in the dynamic component of our model.
Dosages and dosing frequencies can serve as covariates for the effects
of particular medications.

The framework we developed can help establish answers to questions of
interest to clinicians and medical researchers that have been difficult
to assess through simpler models.
In particular, our model can establish the effects of different socio-demographic
or health factors on health outcomes that control for detailed time-varying
adherence to medication by recognizing that medication-taking behaviors
can change over time.
From our framework, we can also estimate the daily improvement in being
adherent to one's medication, controlling for socio-demographic and health
characteristics, but also the likely long-range achievable mean outcomes.
We can also use the model to forecast health outcomes as a function of 
specified patterns of adherence, potentially serving as a tool for 
medical decision-making by clinicians. 
Our approach provides a 
robust framework for understanding the impacts of poor medication adherence 
as clinicians and patients work together to improve their medication treatment.

\newpage

\bibliographystyle{apalike}

\singlespacing

\bibliography{references}

\end{document}